\documentclass[twocolumn]{aastex62}

\graphicspath{{./}{figures/}}

\accepted{November 1, 2018}

\shorttitle{AutoSpec}
\shortauthors{Griffiths}

\begin{document}


\title{\textsc{AutoSpec}: Fast Automated Spectral Extraction Software for IFU Datacubes}

\correspondingauthor{Alex Griffiths}
\email{alex.griffiths@nottingham.ac.uk}

\author[0000-0003-1880-3509]{Alex~Griffiths}
\affil{School of Physics and Astronomy, The University of Nottingham \\
University Park, Nottingham, NG7 2RD, UK}

\author{Christopher~J.~Conselice}
\affil{School of Physics and Astronomy, The University of Nottingham \\
University Park, Nottingham, NG7 2RD, UK}


\begin{abstract}

With the ever growing popularity of integral field unit (IFU) spectroscopy, countless observations are being performed over multiple object systems such as blank fields and galaxy clusters. With this, an increasing amount of time is being spent extracting one dimensional object spectra from large three dimensional datacubes. However, a great deal of information available within these datacubes is overlooked in favor of photometrically based spatial information. Here we present a novel, yet simple approach of optimal source identification, utilizing the wealth of information available within an IFU datacube, rather than relying on ancillary imaging. Through the application of these techniques, we show that we are able to obtain object spectra comparable to deep photometry weighted extractions without the need for ancillary imaging. Further, implementing our custom designed algorithms can improve the signal-to-noise of extracted spectra and successfully deblend sources from nearby contaminants. This will be a critical tool for future IFU observations of blank and deep fields, especially over large areas where automation is necessary. We implement these techniques into the Python based spectral extraction software, \textsc{AutoSpec} which is available via GitHub at: \url{https://github.com/a-griffiths/AutoSpec} and Zenodo at: \url{https://doi.org/10.5281/zenodo.1305848}

\end{abstract}



\section{Introduction} \label{sec:intro}

Spectroscopic analysis of galaxies provides a wealth of information not available from photometric methods.  Most of the advances in astrophysics over the past 100 years have come about due in part to spectroscopy coupled with imaging, and this shows no sign of abating over the next few decades.  Insights provided by spectroscopy include, but are not limited to: radial velocities and redshifts, chemical abundances, internal motions of objects, and the identification of objects along the line-of-sight which can only be seen in absorption, such as Lyman-alpha clouds.

The analysis of a galaxy's content from absorption and emission lines can thus provide an insight into its formation and evolutionary history. The benefits of spectroscopy are ever more prevalent with the introduction of integral field units (IFUs) that can simultaneously obtain spectra over large regions of the sky. IFUs have traditionally been used to determine the internal properties of galaxies, with each optical fiber probing a different physical location within a galaxy.  However, new generation IFUs with large fields of view can now be used to probe galaxy clusters or `blank' fields where in principle, many galaxies are observed within a single IFU pointing.

In current and upcoming eras of astronomy, there is a wealth of information that multi-object IFU observations can provided within these dense, or blank field areas of the universe.  This includes finding galaxies that cannot be seen in the deepest optical imaging \citep{musehudf}, as well as in the study of galaxy clusters \citep[e.g.,][]{clio,museabell}.  Not only does an IFU give information on the radial velocity, and thus membership and physical properties of member galaxies, it also provides information on the background lensed systems.  For example, the accurate identification of multiply imaged galaxies through spectroscopic redshifts provides critical constraints for lensing models. Currently, IFUs are the best, most efficient way to get a complete sample of lensed galaxy redshifts.

An ever increasing amount of scientific research is being conducted with the aid of IFUs such as the multi-unit spectroscopic explorer \citep[MUSE; ][]{muse} and the Gemini Multi-object Spectrograph \citep[GMOS; ][]{gmos}, as well as plans for future instruments on the James Webb Space Telescope (JWST) and the Extremely Large Telescope (ELT). With this, comes the daunting and time consuming process of extracting useful information from the large datacube files produced.

For astronomical images this process is well established, software such as \textsc{SExtractor} \citep{sextractor} is widely used to detect, measure, and classify sources through the creation of photometric catalogs. However, for the analogous process of extracting spectra from three dimensional data, the optimal methodology at this time remains unclear and is typically carried out using various unrefined approaches. 

Many spectroscopic IFU studies are based on the photometric pre-selection of objects, in which catalogs derived from ancillary imaging data, or taken from previous studies are used as a basis of spectral extractions. An alternative comes in the form of software such as the Line Source Detection and Cataloging Tool\footnote{\textsc{LSDCat} available at \url{https://bitbucket.org/Knusper2000/lsdcat}} \citep[\textsc{LSDCat};][]{lsdcat} and the MUSE Line Emission Tracker\footnote{\textsc{MUSELET} is part of \textsc{MPDAF}, documentation available at \url{http://mpdaf.readthedocs.io/en/latest/muselet.html}} \citep[\textsc{MUSELET};][]{mpdaf}. These software packages employ computational techniques to perform blind searches of a datacube in order to identify emission lines. In fact, a combination of photometric pre-selection and blind searches has been found to be favorable \citep[e.g.,][]{musehudf,clio,museabell} to produce source catalogs for spectral extraction. Unfortunately, the optimal method for obtaining one dimensional spectra from an IFU datacube still remains unclear.

The simplest approach is to extract spectra based on fixed apertures, such as is commonly done when measuring galaxy photometry and implemented through source extraction methods and tools such as \textsc{SExtractor} \citep[e.g.,][]{museapp1,museapp2}. An evident drawback to this method is encountered when dealing with more complex sources such as lensing arcs and extended galaxies or emission regions.   

To circumvent some of these issues, an object's morphology can be used when defining spectral extraction regions. IFU studies of galaxy clusters such as \citet{clio} and \citet{museabell} implement the use of \textsc{SExtractor} segmentation maps derived from deep imaging data as a basis of weighted spectral extraction. The work recently carried out on MUSE observations of the Hubble Ultra Deep Field \citep{musehudf} also follows similar extraction methods. It is however, not difficult to imagine situations where this may not be entirely applicable, such as the case where an object has extended emission in wavelength ranges not covered by the available imaging.

Here, we argue that an abundance of spectral information is being overlooked in existing extraction techniques. We present a new method for the identification of the spatial extent of objects directly from IFU datacubes without the need for ancillary imaging or prior knowledge of the sources. The combination of established aperture and segmentation region extraction methods with a simple, but novel custom designed cross-correlation algorithm, can lead to an improvement in spectral signal-to-noise as well as the successful isolation of sources from neighboring contaminants. 

We structure this paper in two main parts; firstly, in Section~\ref{sec:methods} we present our novel technique for the optimal spatial identification of sources directly from a datacube, utilizing the wealth of information available. In Section~\ref{sec:software} we provide an overview of our python based software package \textsc{AutoSpec}, which implements these techniques, along with existing methods for the fast, automated extraction of one dimensional object spectra. We conclude by showing the versatility of the the techniques described in this paper by exploring alternative uses beyond its original design goals.


\section{Optimal Source Identification} \label{sec:methods}

With typical spectral extractions based on circular apertures, or an object's morphology in a particular photometric band, a wealth of information is available within an IFU datacube which is entirely overlooked. Thus, current methods are not ideally suited to spatially identify a source for the purpose of extracting its spectrum. Furthermore, obtaining a sources spectrum from an IFU datacube is a complex process, not to mention the computing power required to handle such large file sizes. With no established methods, we are left to ask questions such as; which spectral pixels (spaxels) correspond to the source, and how best to combine and weight them in order to obtain an optimal 1D spectrum? To answer these questions we present here, a simple but novel technique, combining cross-correlation with continuum extraction to identify and isolate astronomical sources directly from within a datacube itself. 
\vspace{5mm}

\subsection{Cross-correlation}\label{sec:xcor}

Our cross-correlation technique is designed to optimally locate a source from directly within a datacube. In order to calculate the correlation weight, an initial reference spectrum is required. In principle, a spectral template could be used if there is prior knowledge of a sources properties, such and redshift and spectral type. This is however often not the case, so a reference can be obtained via established methods where object masking is best defined either by a circular aperture or morphologically based segmentation region. The first step is to create a truncated cube (subcube) around the source in order to reduce both the processing power and computation time required. From the subcube a spectrum can be obtained via the optimal extraction algorithm \citep{spectrum};
\begin{equation}
f(\lambda)=\frac{\sum_xM_x W_x(D_{x,\lambda}-S_{\lambda}))/V_{x,\lambda}}{\sum_xM_xW_x^2/V_{x,\lambda}}
\label{eq:honre}
\end{equation}
where $f(\lambda)$ is the resultant flux. $M$ is the object mask, $D$ and $V$ are data and variance cubes respectively while $S$ is the sky spectrum. The ideal initial weight image, $W$ is source dependent and can take the form of ancillary board or narrow band imaging. However, if this is not available a `white-light' image created by flatting the datacube along the spectral axis is often sufficient.

Assuming the source is not extended such that Doppler shift gradients are negligible, we employ cross-correlation techniques with zero spectral lag. This provides a measure of similarity between the reference spectrum and the spectrum of each spaxel within the subcube. A two dimensional measure of the cross-correlation strength, $cc(x,y)$ is obtained via the equation;
\begin{equation}
cc(x,y) = f~\star~F(x,y) =\sum_{\lambda}f^*_{\lambda}F_{\lambda}(x,y)
\label{eq:xcor}
\end{equation}
where $cc(x,y)$ is the cross-correlation strength map, $f^*_{\lambda}$ is the complex conjugate of the reference spectrum and $F_{\lambda}(x,y)$ is the subcube spaxels. This cross-correlation technique yields a strength map which details by what degree each spaxel within the subcube correspond to the reference spectrum. A higher value is given to spaxels in which the two spectra are similar (i.e. spaxels which correspond to the source), while lower values are given where spaxels show fewer, or no similarities (i.e. background sky, other objects or contaminants). This method effectively negates any selection effects of morphological analysis via photometrically defined segmentation regions or apertures, while simultaneously providing a weighting scheme for secondary spectral extraction. Further extractions can be performed via Equation~\ref{eq:honre}, using the correlation strength map as a weighting scheme, $W$. In theory, if the source is sufficiently isolated this technique could be applied without any additional masking, however for general cases we have found masking helps to suppress noise and maintain flux conservation. In principle, this weighted extraction technique is a spectroscopic analog of the photometric methods presented in \citet{photoweight}.   

The successfulness of this technique is however limited by the initial reference spectrum used. If a cross-correlation is performed with a reference spectrum that is not a good representation of the object, this technique will provide a less useful map. The main factors which can negatively influence results are noise, and ill defined masks or initial weight schemes. Sources of noise such as neighboring contaminants that are not properly masked out can heavily bias the reference spectrum. For faint objects, where morphologies can not be sufficiently approximated from the white-light or supplementary imaging, we find appropriately sized apertures are best for initial extractions. For more complex sources such as lensing arcs or extended galaxies, morphologically based extractions prove to be most efficient. For sources that are not sufficiently isolated for neighboring objects, we find that the resultant cross-correlation maps can become heavily biased and unreliable, especially when the target source is fainter than nearby contaminants.

\subsection{Isolating and deblending sources}\label{sec:cont}

As previously mentioned, our cross-correlation technique alone is not sufficient to successfully isolate sources from neighboring objects. This issue presents itself when a contaminating object has a similar continuum shape to that of the target source, greatly biasing the resulting cross-correlation strength maps obtained. When this is the case, a continuum subtraction needs to be performed on both the reference spectrum, and each spaxel within the subcube before the correlation strength is measured. This can successfully negate any continuum induced bias to the resultant correlation strength maps obtained, leaving only the spectral emission and absorption features of the source to contribute. 

To obtain an estimate of the continuum, we perform a simple five degree polynomial fit. We use this method for both its speed and simplicity as it is performed on both the reference spectrum, and on each spaxel within the subcube individually. A more robust estimation technique could be employed to include iterative processes and outlier removal, but this would become computationally expensive. Alternatively, a continuum could be estimated from only the reference spectrum and applied to the full subcube, however during our testing we found the resulting strength maps were not as robust.

\begin{figure*}
\includegraphics[width=\textwidth]{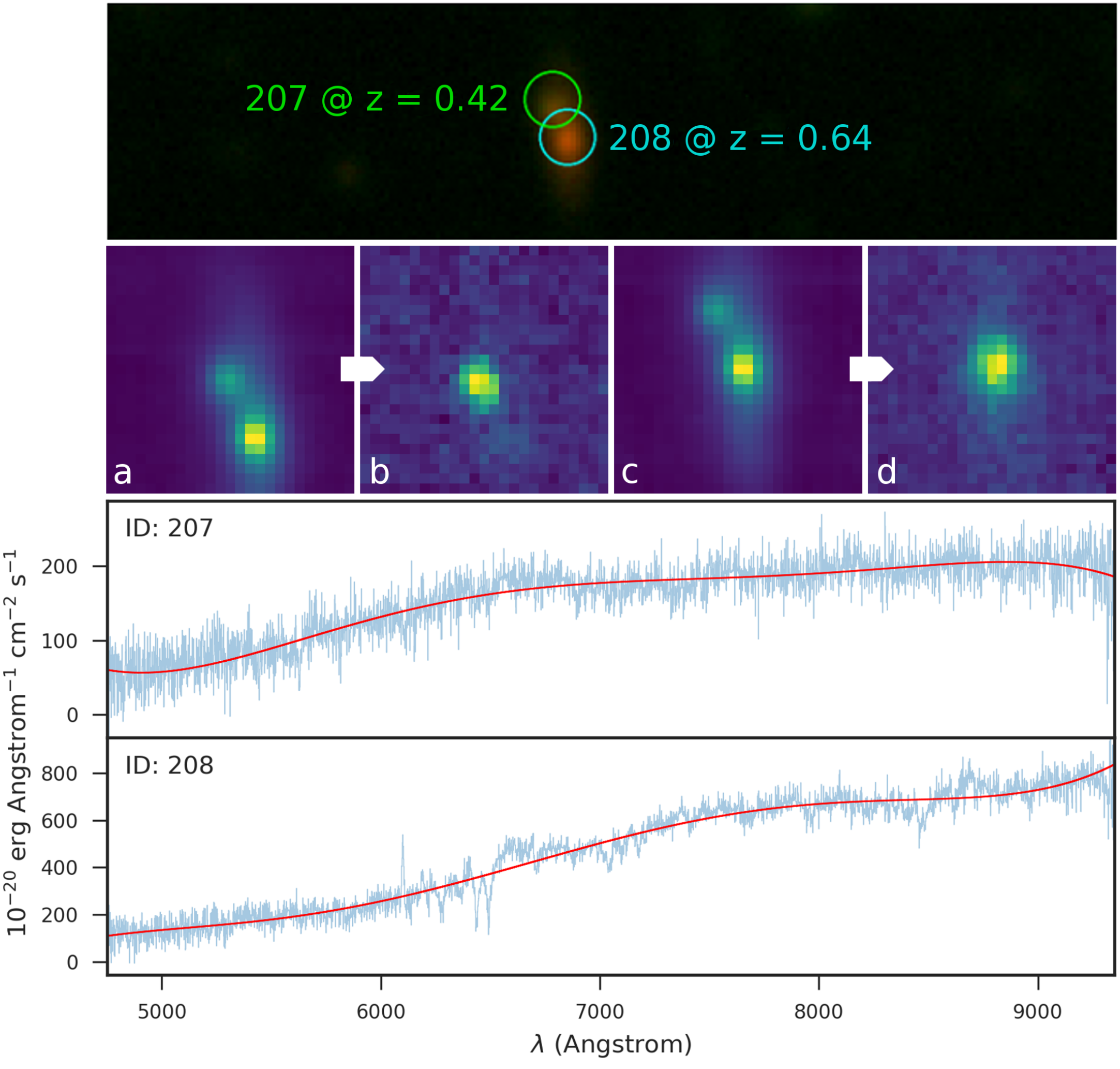}
\caption{We show here the effectiveness of continuum subtraction for a pair of galaxies in close proximity within an observation. Top shows the false color image with source positions overlaid. Bottom two panels show the reference spectrum of each source in blue, and in red we show the continuum estimated using a five degree polynomial fit. Here it can be seen that even though the continuum of the two sources is different, the overall similarity will bias cross-correlation results when the two objects are in this close a proximity. The middle panel a and c show the cross-correlation map without the additional continuum subtraction step for source id 207 and 208 respectively. It can be seen in these images that the contaminating source is also picked up by our cross-correlation methods. b and c show the cross-correlation maps over the same spatial region, this time performed after the additional continuum subtraction routine. Through the comparison of a and b, or c and d, the strength of the cross-correlation algorithm, combined with continuum subtraction is shown to successfully identify a sources spatial distribution and isolate it from nearby contaminants.}
\label{fig:contsub}
\end{figure*}

We find that for objects that are not sufficiently isolated, a combination of cross-correlation and continuum subtraction provides optimal identification of sources within the datacube while simultaneously deblending the source from contaminating objects. A visual example of the effectiveness of this method is shown in Figure~\ref{fig:contsub}. To further improve spectral quality, it is feasible to extend this method to be performed iteratively. In which a spectrum derived in step $i$, can be implemented as reference for iteration $i+1$ in order to obtain more refined cross-correlation strength maps.

\subsection{Signal-to-noise}

To provide some quantification of this method, we compare the signal-to-noise of traditional morphologically and aperture derived spectra to those extracted via our techniques described here. To investigate the signal-to-noise of the spectra we first select various source types with known redshifts. We extract spectra via an appropriately size aperture as well as weighted extractions with masks defined by segmentation regions. We weight these extractions using both the MUSE white-light, and deep g-band imaging. Further, when appropriate we perform additional extractions based on PSF (for stellar spectra) and narrow-band (high redshift) weighting schemes. The narrow-band image is constructed from the datacube, with a width of 100$\AA$ and centered on the Ly-$\alpha$ emission line. 

From these initial extractions, we take the best spectra for each object and use it as a reference for our cross-correlation methods. Extracting a spectrum both before, and after the additional continuum subtraction step. To estimate the signal-to-noise as a function of wavelength we fit a template spectrum to each of the spectra extracted. Template fitting is performed using the Python Spectroscopic Toolkit\footnote{\textsc{PySpecKit} available at \url{https://github.com/pyspeckit/pyspeckit} and \url{https://bitbucket.org/pyspeckit/pyspeckit}} \citep[\textsc{PySpecKit};][]{pyspeckit}. \textsc{PySpecKit} finds the optimal shift and scaling for the given template to accurately match input spectra. We calculate signal-to-noise as a function of wavelength by dividing the shifted, scaled model by the square root of the original spectrum. We find this provides an accurate representation of the noise in order to compare the various extraction methods. 

\begin{figure*}
\includegraphics[width=\textwidth]{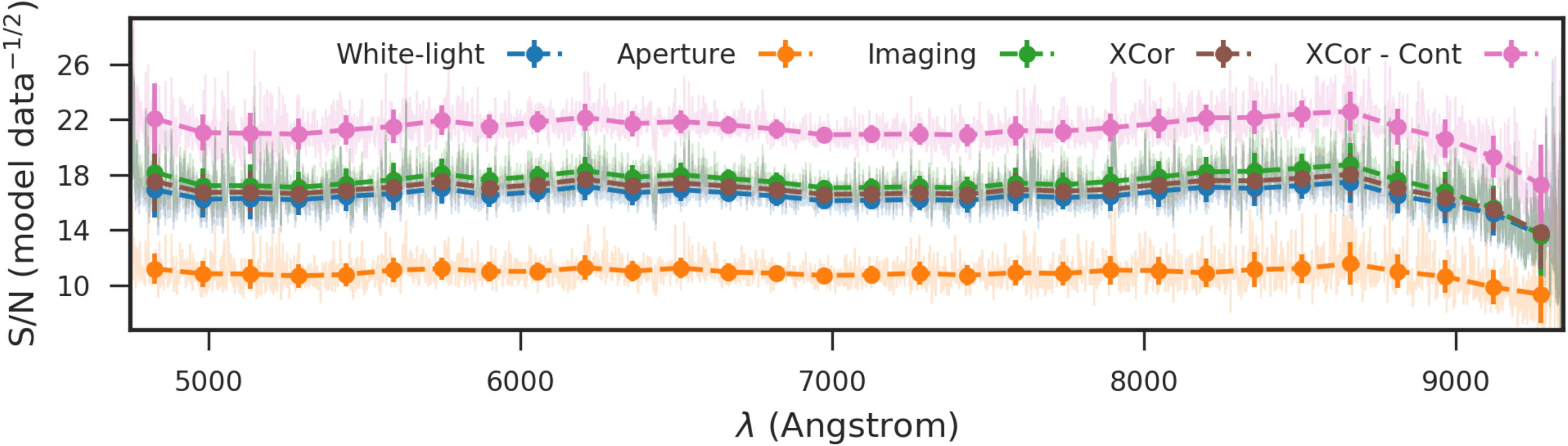}
\caption{We show here the improvement in spectral signal-to-noise through the use of our cross-correlation strength map for an extended lensing arc with various sources of foreground contamination. Additionally, we show the averaged signal-to-noise of this object in Figure~\ref{fig:stn2} as well as the white-light image and cross-correlation strength map in Figure~\ref{fig:examples}.2. In blue and green, we show the signal-to-noise of a spectrum extraction defined via a segmentation region, weighted by the MUSE white-light image and deep g-band imaging respectively. Orange shows the signal-to-noise of a spectrum extracted via a circular aperture region while brown and pink show cross-correlation weighted extractions (with and without the additional continuum subtraction steps respectively). The signal-to-noise is represented as a function of wavelength across the entire range of the IFU datacube, while the dots show the mean data value for bins of $\sim$150\AA. We show that for this object, our cross-correlation technique combined with continuum subtraction improves the signal-to-noise by $\sim$20\% from the next best extraction method (a deep imaging weighted extraction).}
\label{fig:stn1}
\end{figure*}

In Figure~\ref{fig:stn1} we show the spectral signal-to-noise for an example source, an extended lensing arc with nearby foreground contamination (this object can also be seen in Figure~\ref{fig:examples}.2). White-light and deep imaging weighted spectra show an improved signal-to-noise over a circular aperture extraction as can be expected for an extended object. Through the use of our cross-correlation derived strength map alone, we find approximately the same results as imaging weighted extractions. Including the extra continuum subtraction step, we see $\sim$20\% improvement in the spectral signal-to-noise over the next best method. This improvement in signal-to-noise shows that our techniques are able to successfully isolate the source from the foreground contamination, and provided a sufficient weighting scheme for the spectral extraction. Improvements such as this are especially significant when dealing with faint and obscured galaxies, or looking to obtain accurate spectral line measurements.

\begin{figure*}
\includegraphics[width=\textwidth]{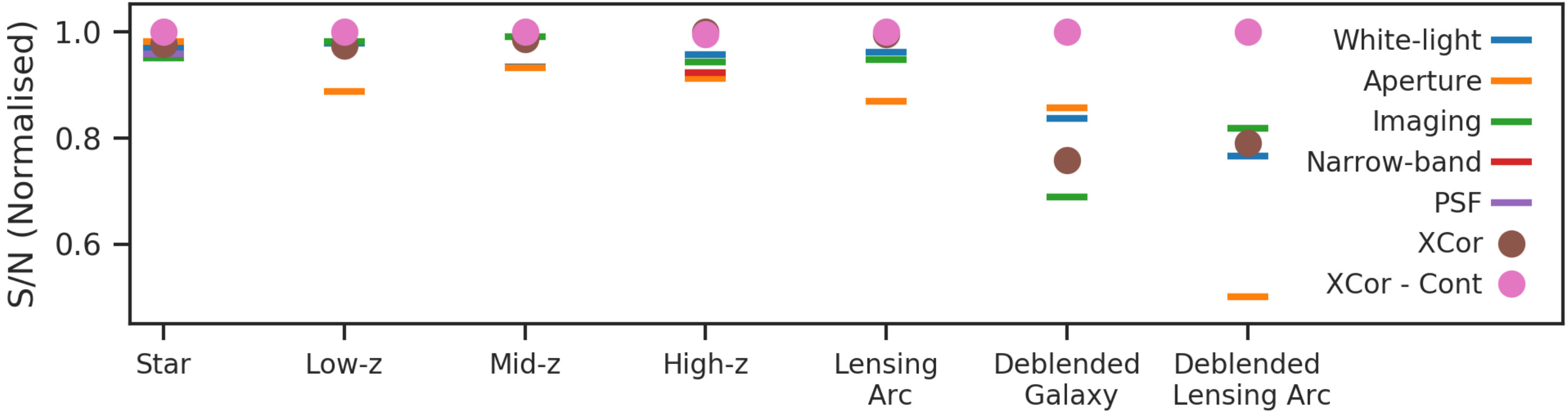}
\caption{We show here the normalized spectral signal-to-noise for a variety of objects and extraction methods. Horizontal lines show traditional weighting schemes: blue and green show spectral extractions weighted by the MUSE white-light image and deep g-band imaging respectively, while orange represents aperture extractions. For the high redshift galaxy we construct a narrow-band image of 100\AA~width, centered on the Ly-$\alpha$ emission line, spectral signal-to-noise derived using this image is shown in red. For stellar spectral extractions we employ an additional psf weighting scheme which we show here in purple. Circular points show the signal-to-noise of spectral extractions weighted via our cross-correlation maps; before (brown) and after (pink) the additional continuum subtraction step. All points shown here represent the median signal-to-noise value across the full wavelength range of the datacube. It can be seen here that in most of the test cases, our cross-correlation methods either improve or are approximately equal to the averaged spectral signal-to-noise of the best traditional extraction method. When it is not, the use of the additional continuum subtraction step helps to improve the resultant signal-to-noise beyond that of traditional methods by an average of $\sim$20\%. Here, the 'Deblended Galaxy' refers to object ID: 208 from Figure~\ref{fig:contsub} and the 'Deblended Lensing Arc' can also be seen in Figures~\ref{fig:stn1} and \ref{fig:examples}.2.}
\label{fig:stn2}
\end{figure*}

We show the results from other object types investigated in Figure~\ref{fig:stn2}. Here, we take the median signal-to-noise value over the entire spectral range to more easily represent the data. We further normalize the signal-to-noise measurements such that the peak value for each source is equal to one. These examples show that for well defined objects such as stars and low redshift galaxies, extractions based on our cross-correlation strength maps provide only a marginal improvement over traditional extraction techniques. However, for more complex sources such as unresolved high redshift galaxies and extended lensing arcs, the implementation of our techniques produces a clear increase in the resulting spectral signal-to-noise. Further, the benefits our technique combined with continuum subtraction is exemplified when considering non-isolated sources. The source labeled as 'Deblended Galaxy' here refers to object ID: 208 from Figure~\ref{fig:contsub}, it can be seen that the use of the cross-correlation alone induces noise from the contaminating galaxy (as described in Section~\ref{sec:cont}). However, when employing the additional continuum subtraction step we find a significant increase in the spectral signal-to-noise over all traditional extraction methods. Similar results can be seen for the 'Deblended Lensing Arc' which is also shown in Figures~\ref{fig:stn1} and \ref{fig:examples}.2.

Again, it is worth mentioning here that any signal-to-noise improvements of the resultant spectrum is highly dependent on the reference used. We find that when the reference is poorly defined, this method is strongly biased by contamination which can result in an overall decrease in signal-to-noise. Similarly, the availability of ancillary imaging data will help define robust reference spectra, and even though in our test cases shown here we find improvements for all objects, for faint sources the white-light image is not always satisfactory to define extractions. A further limitation is the spatial extent of the object, if it is extended such that Doppler shift gradients are non-negligible this technique is not ideal for spectral extractions.  

\section{Software Methods} \label{sec:software}

The python based \textsc{AutoSpec} software we introduce here aims to provide the user with simple, but robust extraction of one dimensional spectra from IFU datacubes using both, existing techniques along with our novel methods described in Section~\ref{sec:methods}. At a minimum, the user is required to supply the software with an IFU datacube along with a catalog of sources to be extracted. A parameter file is supplied which can be used to fine-tune the functionality of the software to the users requirements. \textsc{AutoSpec} makes use of the MUSE Python Data Analysis Framework \citep[MPDAF;][]{mpdaf} for various aspects of source extraction and the construction of the output data files. 

Initial spectral extractions are performed in which the spatial extent is defined either via user defined apertures, or segmentation regions. Segmentation regions can be automatically calculated within the code or can be supplied by the user (see Section~\ref{sec:primary}). With one of the initial spectra defined as a reference, the software performs our custom designed cross-correlation algorithm across a truncated datacube centered on the object (see Section~\ref{sec:secondary}). This cross-correlation algorithm provides a detailed insight into which spaxels correspond to the source in question. This analysis is employed as a unique weighting scheme which can be shown to increase the overall signal-to-noise of the resulting spectra. By performing the additional continuum subtraction step, the software can also successfully deblend sources from neighboring contaminants. In Section~\ref{sec:usage} we provide a brief overview of the required input files as well as output products produced.      

\subsection{Initial extraction} \label{sec:primary}

For the first step in the extraction procedure \textsc{AutoSpec} iterates through each source in the input catalog and creates a subcube from the supplied IFU datacube. The subcube is centered on the source co-ordinates with its extent defined by the user. The creation of the subcube is a necessary step in improving computational memory usage as well as processing time.

To provide the user with as much flexibility as possible the \textsc{AutoSpec} software automatically extracts initial object spectra based an individual, or multiple user defined apertures, weight images and segmentation maps. Firstly, aperture spectra are calculated from within circular regions with no additional weighting applied. Secondly, the software will use all user supplied images to derive a segmentation region using Source Extractor, parameter files for which can be supplied by the user if required. It is also possible to perform segmentation region extraction without additional data, however, by supplying ancillary imaging data or segmentation maps, extraction regions can be more accurately estimated. This is especially important for faint sources that are unlikely to be detected directly from the datacubes white-light image. For each additional image supplied, as well as the MUSE white-light image, a weighted spectra will be calculated using Equation~\ref{eq:honre}. Further, if the user has access to existing segmentation maps these can be supplied in place of, or in addition to those calculated within the software. 

\subsection{Improving spectral quality} \label{sec:secondary}

To make use of the wealth of information available within the datacube, we provide the user with the option to implement our cross-correlation and continuum subtraction algorithms in order to deblend sources, and perform secondary spectral extractions if required. 

If this step is to be undertaken, the user is required to define one of the preliminary spectra (obtained as described in Section~\ref{sec:primary}) as a reference, this can be done on a source-to-source, or per run basis. The software performs our cross-correlation algorithm across the full subcube using Equation~\ref{eq:xcor} and the methods described in Section~\ref{sec:xcor}. The masks used to produce the reference spectrum are also used in this step. This additional masking is not always necessary however we find in most cases it helps to negate sources of noise and improve flux conservation in these secondary extractions.  This analysis yields a weight map, providing a detailed description of the extent of the source within the datacube itself. This weight map is then used as the basis for a secondary source extraction. 

\begin{figure*}
\includegraphics[width=\textwidth]{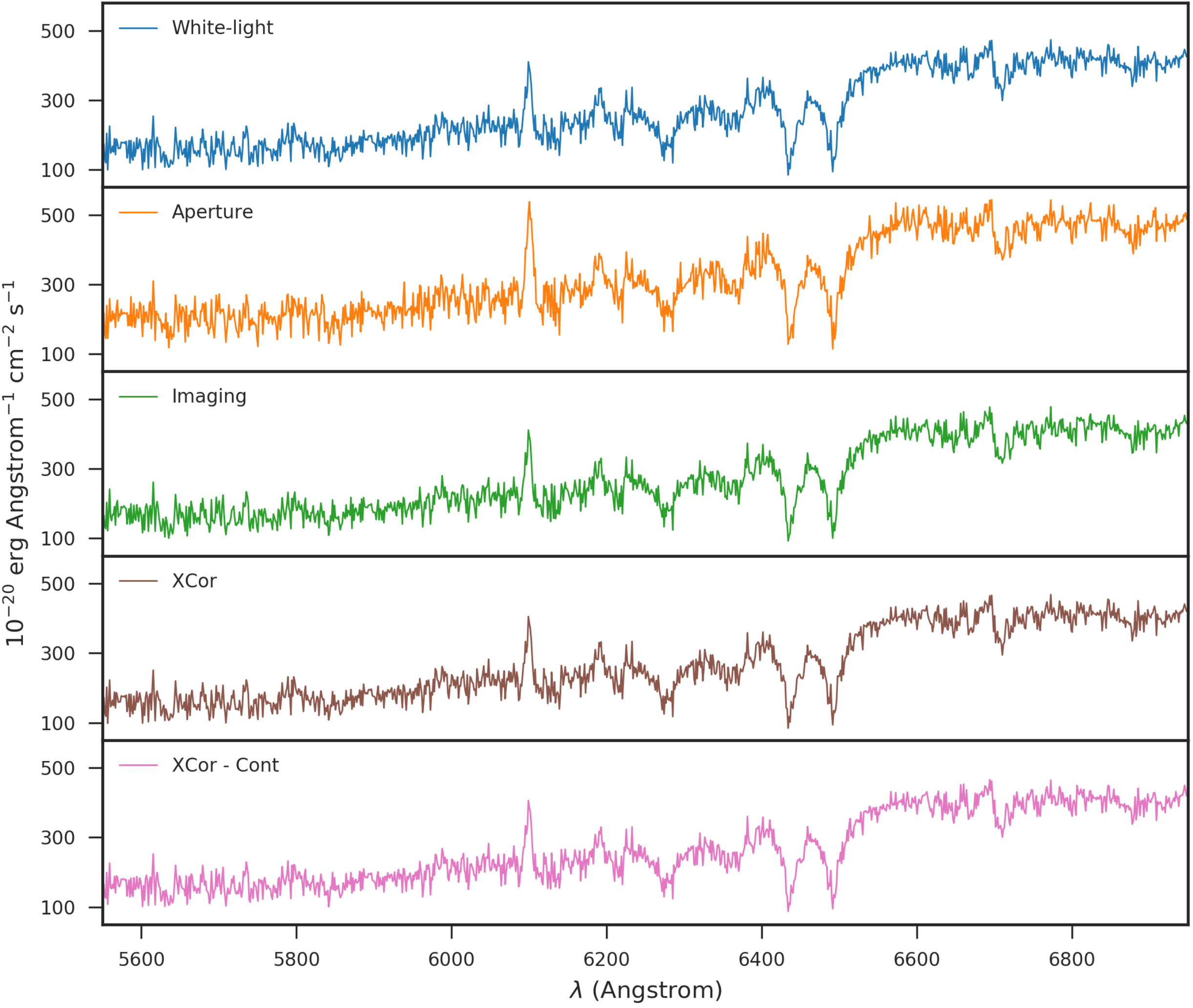}
\caption{Example galaxy spectra extracted via the various methods available within \textsc{AutoSpec}. In blue, orange and green we show spectra extracted using traditional methods. Orange shows an extraction defined by a circular aperture while blue and green are masked using segmentation regions generated within the software and weighted by the MUSE white-light, and deep g-band imaging respectively. The g-band weighted spectra is used as a reference in order to define a cross-correlation weight map both before and after continuum subtraction and used to weight spectra shown in brown and pink respectively. For cross-correlation extractions, \textsc{AutoSpec} uses the same masks as used by the reference spectrum in order to reduce sources of noise and maintain flux conservation.}
\label{fig:spectra}
\end{figure*}

If the subcube is likely to be contaminated by neighboring objects, the user can also choose to perform the additional continuum subtraction step here. Following the methods described in Section~\ref{sec:cont} the target source can be isolated from a neighboring objects. While subtracting the continuum from the subcube increases the processing time of each source extraction, we find that the resulting spectral quality can be greatly improved (see Figure~\ref{fig:stn2}). Following this step, \textsc{AutoSpec} will produce an additional secondary spectral extraction. In Figure~\ref{fig:spectra}, we show examples of the spectra extracted for a single object in a run of the \textsc{AutoSpec} software. In this case, the software is supplied with a single aperture and an additional image. While it is difficult to see by eye any noticeable improvements in the spectral quality, we note that flux conservation is maintained through all spectral extractions methods undertaken by \textsc{AutoSpec}.

\vspace{5mm}

\subsection{Using the software} \label{sec:usage}

The software has been designed to be as user friendly as possible. The user is required to supply the IFU datacube along with a catalog of the central positions (RA and DEC) of sources. The catalog can be supplied in one of two different formats, if settings are provided on a per run basis the first three columns of the catalog are required to be in the format of: Source ID (integer), right ascension (degrees) and declination (degrees). This is compatible with a wide variety of existing catalogs, including those produced by \textsc{MUSELET} which can be implemented directly to \textsc{AutoSpec}. Alternatively, if the user wants to define extraction parameters on a source-to-source basis, they are required to provide two additional columns of data: extraction size of subcube (in arcseconds) and a reference spectrum label (either aperture or weight image).

The user can supply additional images from which the segmentation region can be defined and weighted extractions will be undertaken. The user can also directly supply \textsc{SExtractor} segmentation maps derived independently of \textsc{AutoSpec}. Our software runs through the command line interface via the python environment and all user settings can be configured via the provided parameter and catalog files. The \textsc{AutoSpec} software and detailed usage instructions are available on GitHub\footnote{\url{https://github.com/a-griffiths/AutoSpec}}.

We test our software on a standard research computer (Intel i3-6100 3.70GHz CPU with 8GB of ram). After a one time initial set up procedure per run (which will vary depending on the size of the datacube and number of additional images supplied), source extraction typically takes 3-4 seconds per object. This includes 3 aperture extractions, 3 image weighted extractions (including white-light), as well as calculating and extracting cross-correlation weighted spectra before and after the additional continuum subtraction step. Effectively processing catalogs of hundreds, or thousands of objects in a very short time spans.

\subsubsection{Output}

For each source successfully extracted, the user is presented with a fits format file, the contents of which can be customized according to the users preferences. Additionally, for the ease of the user, \textsc{AutoSpec} can output jpg files showing the generated masks, cross-correlation weight maps and spectra obtained for each object. 


\section{Alternative Uses}

The development of the techniques and software as detailed in this paper are motivated by work on lensing clusters where the identification and extraction of spectra from a MUSE IFU datacube proved to be a laborious process. However, we show here that their application is not limited to a particular type of observation or instrument.

For IFU observations of single galaxies such as those obtained in the Mapping Nearby Galaxies at APO survey \citep[MaNGA;][]{manga}, our cross-correlation techniques are able to spatially identify regions with common spectral features. The produced cross-correlation maps may also help to identify the spatial extent of particular galactic components. Additionally, we suggest that a combination of the cross-correlation and continuum extraction techniques as detailed in this paper are ideal for the identification of multiple-images to constrain strong lensing models, this however would require significant computing power to be run across large datacubes.  

\begin{figure}
\centering
\includegraphics[width=1\linewidth]{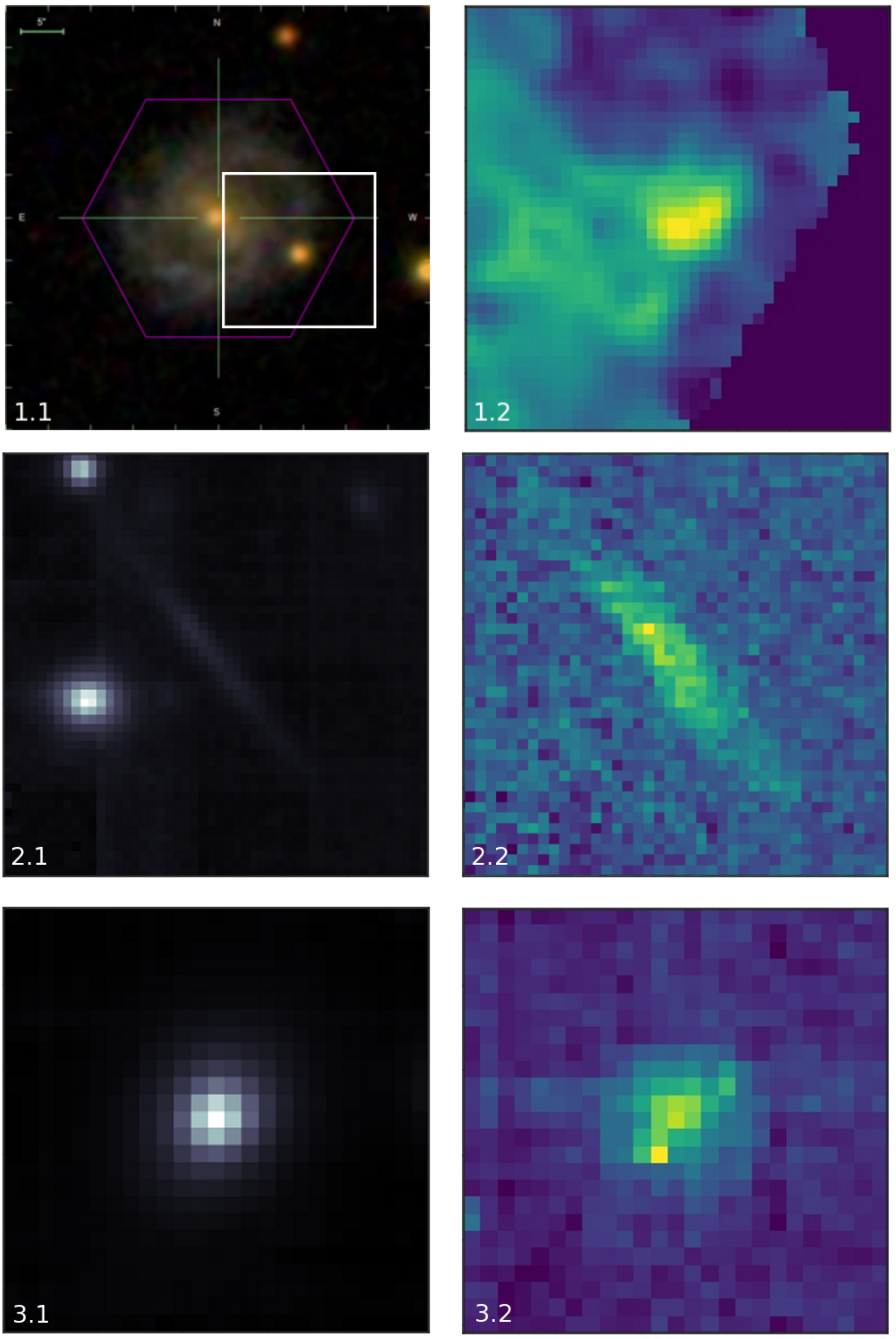}
\caption{Here we show 3 examples demonstrating the power of our source identification methods. Top (1.1) is an SDSS image of a galaxy observed in the MaNGA survey \citep{manga}, the overlaid white box shows the extraction region and top-right (1.2) shows the cross-correlation map. Middle (2.1 and 2.2) show the MUSE white-light image and cross-correlation strength map of a lensing arc detected in the CLIO cluster \citep{clio}. Bottom (3.1 and 3.2), shows a white-light image of a Quasar and the corresponding cross-correlation map from MUSE observations of Quasar Field SDSS J1422-00 \citep{quasar}.}
\label{fig:examples}
\end{figure}

In Figure~\ref{fig:examples}, we show the versatility of our cross-correlation method using various datacubes and observations.

\section{Summary} \label{sec:summary}

We find that by utilizing the wealth of information available within IFU datacubes, we are able to isolate sources and can obtain increased signal-to-noise spectra. Our cross-correlation algorithm paired with continuum subtraction performs consistently well at deblending sources and providing a unbiased weighting scheme for spectral extractions.

As these techniques are designed for the extraction of a single, one dimensional spectrum per object, its usefulness is limited to observations in which sources do not subtend large areas of the sky such that Doppler shift gradients are negligible. Thus, it is best employed for cluster or field studies where these velocity gradients will have minimal effect. As the production of this software was motivated by the work carried out in \citet{clio}, we find it is particularly useful for observations of lensing clusters where it is able to successfully identify and extract the spectra of both cluster and background galaxies, as well as extended lensing arcs. However, we have shown that its application is not limited to these types of observations. 

We provide a simple to use tool for the spectral extraction of small or large catalogs of objects with minimized set-up and run time. While this software has been designed with MUSE observations in mind, it is applicable to any IFU data, provided it can be read by the MPDAF python package. We make this software available under a BSD 3-Clause License via Zenodo \citep{zenodo} and GitHub at: \url{https://github.com/a-griffiths/AutoSpec}.


\acknowledgments
\vspace{5mm}

We thank the anonymous referee for the thorough review and helpful comments and suggestions, which significantly contributed to the improvement of the manuscript. We acknowledge the \textsc{MPDAF} team for providing a useful framework for our software as well as their valuable assistance. This work was supported by the Science and Technology Facilities Council.

\vspace{5mm} 

\facilities{VLT:Yepun (MUSE)}

\software{\textsc{AutoSpec} \citep{zenodo}, \textsc{LSDCat} \citep{lsdcat}, \textsc{MPDAF} \citep{mpdaf}, \textsc{PySpecKit} \citep{pyspeckit}, \textsc{SExtractor} \citep{sextractor}}
          



\end{document}